\documentclass[preprint2]{aastex}
\usepackage{graphicx}
\usepackage{amssymb}

\shorttitle{Viscous evolution of the Enceladus torus}

\begin{document}
\bibliographystyle{icarus}

\title{Saturn in hot water: viscous evolution of the Enceladus torus}
\author{Alison J. Farmer}
\affil{Harvard-Smithsonian Center for Astrophysics, Cambridge, MA 02138}
\email{afarmer@cfa.harvard.edu}

\begin{abstract}
The detection of outgassing water vapor from Enceladus is one of the
great breakthroughs of the \emph{Cassini} mission. The fate of this water once ionized has been widely studied; here we investigate the effects
of purely neutral-neutral interactions within the Enceladus torus.
We find that, thanks in part to the polar nature of the water molecule,
a cold ($\sim 180$ K) neutral torus would undergo rapid viscous heating and spread to the
extent of the observed hydroxyl cloud, before plasma effects become important.  We investigate the physics behind the
spreading of the torus, paying particular attention to the competition between heating and
rotational line cooling. A steady-state torus model is constructed, and it is demonstrated that the torus will be observable in the millimeter band with the upcoming \textit{Herschel} satellite. The relative strength of rotational lines could be used to distinguish between physical models for the neutral cloud.
\end{abstract}


\section{Introduction}

Undoubtedly one of the major discoveries of the \emph{Cassini} mission was the plume of water vapor and dust emanating from Saturn's moon Enceladus \citep{hansen.2006,porco.2006,waite.2006}. In addition to opening up many lines of research as to the origin of the vents and their source of heat \citep[e.g.][]{roberts.2008,meyer.2007,halevy.2008}, the discovery solved a hitherto outstanding mystery of the Saturn system: since 1993 it has been known that there exists a large population of hydroxyl radicals in the inner magnetosphere \citep{shemansky.1993, jurac.2002}, centered near the orbit of Enceladus at $4 R_S$ ($1 R_S=6.0  \times 10^9$ cm) and extending to $R \sim 10 R_S$. In order to populate and maintain the OH torus, it was estimated that $\dot N_W \sim 10^{28}$ molecules per second of water needed to be injected into the magnetosphere \citep{jurac.2005} from a then unknown source. \cite{hansen.2006} estimate that the release of water from Enceladus is sufficient ($\dot N_W > 5 \times 10^{27} {\rm ~s}^{-1}$) for the small moon to be the source of the observed OH torus.

The ``tiger stripe'' regions from which the water plume emerges \citep{spitale.2007} possess inferred temperatures as high as 180K \citep{spencer.2006}. Recently, a hotspot temperature of 180 K was measured, again by the CIRS instrument (NASA news release 2008-050). The gravitational pull of  Enceladus (escape velocity $0.2 {\rm ~km~s}^{-1}$) does little to slow the outflow of gas at this temperature. Free expansion of water vapor at 180K  from the surface of Enceladus and subsequent intersection of the jet with previously emitted vapor will result initially in a narrow ($\Delta R \simeq 0.4 R_S$) torus of vapor with characteristic temperature $\lesssim 180$K, i.e. velocity dispersion $u \lesssim 0.5 {\rm ~km~s}^{-1}$ \citep{johnson.2006}. 

The lifetime of the neutral water is limited by ionization and dissociation processes whose rates are listed in Table \ref{tab:proc}.  \cite{johnson.2006} demonstrated that radial spreading of the water torus can occur via one of these processes: charge exchange with corotating ions. In this paper we investigate a different spreading mechanism. From Table \ref{tab:proc} it is clear that a neutral molecule does not ``feel'' the influence of the magnetospheric plasma or solar UV during a time $\tau \gtrsim 6 \times 10^6$ s after production. We shall demonstrate that  this is sufficient time for the neutral water molecules to interact with \emph{each other} many times, and for the torus to heat and spread to the extent of the observed OH torus.

The dynamics of moon-generated tori have been considered by \cite{smyth.1993} in the context of the atomic hydrogen tori of Titan and Triton. \cite{marconi.2003} and \cite{decker.1994} investigated the viscous heating of the Triton torus via monte carlo simulations. The Enceladus torus is unique in consisting of nearly pure water vapor \citep{waite.2006}. The particular properties of water as a polar molecule make the Enceladus water torus a fascinating physical and chemical system. Comets are another astrophysical system in which almost pure water is observed, and much of the work on the spectrum and collisions of water vapor has been done in this context \citep[e.g.][]{bensch.2004,buffa.2000}. We shall make use of these results for the cometary case.
 Global simulations of the water torus around Saturn, including all species of neutrals and ions, have been carried out by \cite{jurac.2002} and \cite{jurac.2005,jurac.2007}. In these studies, no significant viscous heating was found. We discuss the reasons for this discrepancy in Section \ref{sec:discuss}.
 
In Section \ref{sec:xs} we calculate the cross section for collisions between water molecules, and estimate the collision rate in the water cloud. 
The nature of viscous evolution at low collision rates is discussed in Section \ref{sec:stir}, where we also  investigate the impact of collisional inelasticity upon both spreading and heating rates.
Section \ref{sec:cooling} is devoted to the evaluation of the inelasticity of neutral water collisions in the Enceladus torus, and then in Section \ref{sec:spread} we find the degree of viscous evolution permitted before the neutrals are lost to charge exchange. The steady-state column density profile of the heated torus is estimated in Section \ref{sec:sstate}. In Section \ref{sec:discuss} we discuss the consequences of the neutral-neutral interaction and make comparisons with previous studies. The suitability of the upcoming \textit{Herschel} mission for observing the neutral water is considered in Section \ref{sec:obs}.

\begin{table}[htdp]
\caption{Rates of external (non neutral-neutral) processes. Charge exchange and electron impact ionization rates are applicable at radial location of Enceladus ($4 R_S$), assuming ions and neutrals occupy the same volume.  Notes: (a) from \cite{johnson.2006}; (b) from \cite{burger.2007}}
\begin{center}
\begin{tabular}{l|r}
Process & Rate (s$^{-1}$)\\
\hline
Photodissociation & $1.5 \times 10^{-7}$ (b)\\
Photoionization & $7 \times 10^{-9}$ (b)\\
Charge exchange & $1.7 \times 10^{-7}$ (a)\\
Electron impact ionization & $1.6 \times 10^{-8}$ (b)\\
Keplerian orbital & $5 \times 10^{-5}$~~~~~~\\
 frequency at Enceladus,  $\Omega_K$&\\
\end{tabular}
\end{center}
\label{tab:proc}
\end{table}%

\section{Collision cross section}
\label{sec:xs}

The water molecule possesses a permanent dipole moment $\mu = 1.85 \rm{~D} = 1.85 \times 10^{-18} {\rm ~esu~cm}$, which permits long-range interactions between molecules and leads to interaction cross sections substantially larger than its nominal ($\sim 5 {\rm ~\AA}^2$) size. We expect the collision cross section $\sigma$ appropriate for water-water scattering to scale as follows. A  substantial change of direction occurs when the dipole interaction energy is of order the molecular kinetic energy at temperature $T$, i.e. 
\begin{equation}
\frac{3 k T}{2} \sim \frac{\mu^2}{r^3} \, ,
\end{equation}
where $r$ is the separation between molecules. We then use $\sigma \sim \pi r^2$ to obtain 
\begin{equation}
\sigma \sim  \pi \left(\frac{2 \mu^2}{3 k T}\right)^{2/3}\, ,
\label{eq:water_coll}
\end{equation}
which for water at 180 K gives $\sigma \simeq 54 {\rm ~\AA}^2$. 
For a more accurate value, appropriate for use in our calculations involving viscous spreading, we employ the results of \cite{teske.2005}, who measured experimentally the viscosity of water vapor at temperatures 300-450K and a range of densities. We use the standard relation for kinematic viscosity $\nu$,
\begin{equation}
\nu = \frac{u}{3 n \sigma}\, ,
\label{eq:high_visc}
\end{equation}
where $n$ is the number density and $u$ the thermal velocity, to express their results in terms of a collision cross section. Their resulting $\sigma(T)$ closely follows the power law temperature dependence predicted in Eq.  \ref{eq:water_coll}. Extrapolating to 180K yields $\sigma(180)  = 80 {\rm ~\AA}^2$. Hereafter we use
\begin{equation}
\sigma_{\rm coll} =  80 \left(\frac{T}{180 {\rm K}}\right)^{-2/3}  {\rm ~\AA}^2
\label{eq:sigmaT}
\end{equation}
for the water-water scattering cross-section.

To assess the importance of neutral-neutral interactions in the Enceladus torus, we perform some order of magnitude calculations. Adopting a production rate of water  $\dot N_W \simeq 1 \times 10^{28} {\rm ~s}^{-1}$ \citep{jurac.2005}, and assuming initially that the molecules do not interact with each other, the total number of water molecules in the torus would be
\begin{equation}
N_W \sim \dot N_W \tau_{\rm ce}
\end{equation}
where $\tau_{ce}\simeq 6 \times 10^6 {\rm ~s}$ is the lifetime to charge exchange with water group ions (see Table \ref{tab:proc}). 
A torus of water at 180K at the orbit of Enceladus has extent $H \sim 2 v/\Omega_K \sim 0.4 R_S$ in the vertical and radial directions. We assume that the torus is axisymmetric; the azimuthal spreading time $\tau_\phi \sim 8 \pi R_S/u \simeq 3 \times 10^6$ s is not dramatically different from the lifetime of the individual molecules to charge exchange, but we believe axisymmetry to be a reasonable approximation for the purposes of our investigation. Departures from axisymmetry will be considered elsewhere. Dividing the total number of molecules by the volume of the torus gives a density $n \sim 7 \times 10^4 {\rm~cm}^{-3}$. We now allow ourselves to consider collisions between neutrals.  At this density, the water-water collision timescale is
\begin{equation}
\tau_{\rm coll}  \sim (n \sigma_{\rm coll} u)^{-1} \sim 4 \times 10^4{\rm~s} \, ,
\label{eq:tcoll}
\end{equation}
where we used Eq. \ref{eq:sigmaT} to obtain $\sigma_{\rm coll}$. The neutral water molecules can thus collide $\tau_{\rm ce}/\tau_{\rm coll} \sim 150$ times before suffering a charge exchange collision. The rest of this study is devoted to determining the consequences of these purely neutral interactions that occur before the water feels the presence of the magnetospheric plasma. Molecular collisions will lead to viscous spreading and heating of the torus, effects which are particularly pronounced when there is fewer than one collision per molecule per orbit ($\Omega_K \tau_{\rm coll} \gtrsim 1$). For the collision timescale derived above,  $\Omega_K \tau_{\rm coll} = 2$; as the torus expands its density will fall and the time between collisions will increase, so we expect the torus always to be in the low collision rate regime. We explore this regime in the next section.

\section{Evolution at low collision rate}
\label{sec:stir}
The action of molecular viscosity in a shearing torus is to transport angular momentum outwards and to dissipate gravitational energy as heat while the center of mass of the torus moves to lower gravitational potential. The viscous behavior of such a system depends critically on the number of collisions per keplerian orbit per molecule, i.e. on the parameter $\Omega_K \tau_{\rm coll}$. When there are many collisions per orbit, the standard formula for viscosity Eq. \ref{eq:high_visc} applies. In this regime a lower collision rate results in a higher viscosity, due to the reduction in the molecular mean free path. When the collision rate is low enough, however, ($\Omega_K \tau_{\rm coll} \gtrsim 1$) the opposite applies. Particles are each ``trapped'' within a radial annulus of width $\sim 2eR$, where $e$ is the orbital eccentricity. In this regime, a decreased collision rate means less contact between radial shells and hence lower viscosity. Thus when $\Omega_K \tau_{\rm coll} \simeq 1$ (as in the Enceladus torus) the viscosity is in a sense maximal. The viscous heating rate per molecule is accordingly also maximal and, when there is no cooling, corresponds to an exponential growth of temperature. This behavior can be understood in terms of collisional reorientation of the particles' anisotropic velocity ellipsoid in the low collision regime \citep[see e.g.][]{goldreich.2004}.

The temperature evolution of the torus depends critically on whether the viscously dissipated heat can be radiated away or otherwise removed from translational kinetic energy. Due to their internal degrees of freedom, molecular collisions are frequently inelastic. To determine the fate of the Enceladus torus, we must know the extent to which inelasticity affects viscous evolution in the low collision rate regime. 
We quantify the problem as follows. Our derivation is based on \cite{narayan.1994}, who work in the shearing sheet approximation and take moments of the Boltzmann equation for the gas particles within the sheet. The quantities we wish to study are the second velocity moments: the total velocity dispersion $u^2$ and its components along each axis, $u^2= u_r^2+u_\phi^2+u_z^2$; and $\overline{u_r u_\phi}$, the shear stress. \cite{narayan.1994} assume no spatial gradients in these second moments, an approximation only strictly valid when $u \ll W_r \Omega_K$, where $W_r$ is the radial width of our torus. Collisions are taken to be isotropic; this  is acceptable when we use a collision cross section derived from measurements of viscosity.

\cite{narayan.1994} considered the steady state; we retain time variation (first term in their eq. 3.1.11) with an additional parameter $\zeta$ to represent heating; $\zeta$ is the fractional increase of the second moments per collision. The time evolution of these moments is then
\begin{equation}
\frac{d u^2}{dt}= \frac{\zeta}{\tau_{\rm coll}} u^2 \; ; \; \frac{d (\overline{u_r u_\phi})}{dt}= \frac{\zeta}{\tau_{\rm coll}} (\overline{u_r u_\phi})\, .
\end{equation}
As in \cite{narayan.1994}, the fractional inelasticity per collision is denoted $\xi$. From their Eq. 3.6.3-3.6.6 and specializing to a Keplerian disk, we obtain the following equations for the four non-vanishing second moments:
\begin{eqnarray}
3(1+\zeta) u_r^2-12 \Omega_K \tau_{\rm coll}  \overline{u_r u_\phi} - (1-\xi)u^2 &=& 0 \, , \\ 
3(1+\zeta) u_\phi^2+3 \Omega_K \tau_{\rm coll}  \overline{u_r u_\phi} - (1-\xi)u^2 &=& 0\, ,\\ 
3(1+\zeta) u_z^2- (1-\xi)u^2 &=& 0\, ,\\ 
(1+\zeta)  \overline{u_r u_\phi}- \Omega_K \tau_{\rm coll} \frac{u_r^2}{2} - 2 \Omega_K \tau_{\rm coll} u_\phi^2 &=&0 \, .
\end{eqnarray}
Solving the equations under the assumption $\Omega_K \tau \gg 1$ yields
\begin{equation}
\zeta=(3-11 \xi)/8\, ,
\label{eq:heat}
\end{equation}
\begin{equation}
u_r^2=(20/33)u^2\, ,
\label{eq:ur}
\end{equation}
\begin{equation}
u_\phi^2=(5/33) u^2\, ,
\end{equation}
\begin{equation}
u_z^2=(8/33)u^2\, ,
\label{eq:uz}
\end{equation}
\begin{equation}
\overline{u_x u_y} = \frac{(1+\zeta)}{11 \Omega_K \tau} u^2 \, .
\end{equation}
We see from Eq. \ref{eq:heat} that $\zeta>0$, i.e. the velocity dispersion increases with time so long as 
\begin{equation}
\xi <\xi_{\rm lim} = 3/11 \, .
\label{eq:elimit}
\end{equation}
Greater inelasticity leads to cooling. In Section \ref{sec:cooling} we determine the inelasticity of water-water collisions and find that according to the above condition, the torus will increase in temperature.

We can also obtain immediately an expression for the viscosity,
\begin{equation}
\nu = \frac{-\overline{u_xu_y}}{\partial v_\phi /\partial r} = \frac{(1-\xi)}{12} \frac{u^2}{\Omega_K^2 \tau} \, ,
\label{eq:visc}
\end{equation}
where the rate of shear $\partial v_\phi/\partial r = 3 \Omega_K/2$ in the shearing sheet approximation. We see that finite inelasticity $\xi$ also decreases the viscosity, though not as severely as it does the heating rate. Thus there can be substantial spreading without heating.

\section{Cooling}
\label{sec:cooling}
\subsection{Internal modes of the water molecule}

At the energies considered in this paper, molecular rotation is the most important internal mode. The lowest vibrational mode corresponds to energy $\sim 0.2$ eV. Dissociation of the molecule requires 5.1 eV, while ionization needs 12.6 eV. 

The water molecule is an asymmetric top with principal moments of inertia 1.0, 2.9 and $1.9 \times 10^{-40} {\rm ~g~cm}^2$. The lowest rotational transitions are listed in Table \ref{tab:rot}; the molecule effectively possesses two ground states, corresponding to spin 1 (ortho) and spin 0 (para) wavefunctions. 

\begin{table}
\caption{The lowest rotational transitions of water, from \citet{buffa.2000}}
\begin{center}
\begin{tabular}{c|c|c}

Transition &$1_{11}  -  0_{00}$ & $1_{10} -  1_{01}$ \\
 $J_{K_+ K_-} - J'_{K'_+ K_-'} $ & (para) & (ortho)\\
\hline
Wavelength (mm) &0.27 & 0.54 \\
Energy (eV) &0.005 & 0.0025 \\
A (s$^{-1}$) & $5.6 \times 10^{-2}$ & $3.5 \times 10^{-3}$
\end{tabular}
\end{center}
\label{tab:rot}
\end{table}%

The population of rotational levels is far from thermal equilibrium. Comparing the spontaneous emission coefficients from Table \ref{tab:rot} to the collision time estimated in Eq. \ref{eq:tcoll} reveals $A \tau_{\rm coll} \gtrsim 100$ (this remains true for spontaneous emission from higher rotational levels). Thus any rotational modes excited by a collision will have decayed radiatively to the ground state before the next collision. Energy that goes into molecular rotation is effectively lost to the system and constitutes cooling. Further, when considering collisional excitation we need only include transitions from the ground states.

\subsubsection{Other excitation mechanisms}

Internal modes may be excited by other processes, such as solar radiation or microwave background pumping, or collisions with electrons (we neglect collisions with ions here because we consider these to be a loss mechanism for the neutral water). Solar pumping results in an excitation rate $\sim 10^{-6} {\rm ~s}^{-1}$  \citep[from values for comets, in][]{crovisier.1984,bensch.2004}, still well below the spontaneous emission rate. The microwave background produces an excitation rate $\sim 10^{-7} {\rm ~s}^{-1}$.

The maximum density of electrons in the magnetosphere is $n_e \sim 100 {\rm cm}^{-3}$ \citep{wahlund.2005}. For any reasonable interaction speed and cross section, the rate of electron impacts is much less than $A \gtrsim 10^{-3} {\rm s}^{-1}$. We conclude therefore that the water molecules are essentially always to be found in their ground state, and that rotational energy cannot be converted into translational; rather it can only flow in the other direction, via collisions.

\subsection{Transitions between internal modes: quantum view}

The typical kinetic energy of a molecule in the cold torus at 180K is $E_{\rm trans} = 3kT/2 \simeq 0.023$ eV, not much larger than the energy spacing between the lowest rotational levels. At these lowest temperatures, we cannot neglect quantum mechanical effects.

The permanent dipole moment of the water molecule makes possible long range interactions that involve transitions between internal modes. In particular, the dipole-dipole interaction potential leads to the same selection rules as for dipole allowed radiative transitions \citep[e.g.][]{buffa.2000}. In the case of rotational transitions, this corresponds to $\Delta J = \pm 1, 0$ (the parity of the wavefunction must change, but the total spin must not).

Resonant exchange of a quantum of rotational energy is the most common type of interaction \citep{mason.1962}; since in our case molecules are in the ground state when they collide, these do not concern us.
Exact quantum mechanical calculations for the rate of translational to rotational energy are unavailable.
Instead we make a ``worst case scenario'' estimate of the maximum amount of energy lost per collision, if in every collision a rotation were excited. At 180K, the maximum energy loss per collision that still permits viscous heating is 0.006 eV (3/11 of the thermal energy). 

The ratio of ortho to para water is not known a priori. For some comets (temperature $\sim 100$ K) it has been measured  at $\simeq 3$ \citep{mumma.1987,crovisier.1997}: we adopt this value. The average energy loss to rotation per collision could then be at most  $0.005 \times 0.75+0.0025\times 0.25 = 0.0044$ eV, i.e. heating would still occur.  In fact, \cite{mason.1962} estimate that only once in every 4 collisions does transfer between translational and rotational energy occur, making the the average energy lost per collision of order 0.001 eV.

\subsection{Transitions between internal modes: classical view}

From a classical point of view, the water molecule consists of frictionless atoms of oxygen and hydrogen (two). Only individual atoms within molecules are considered to collide during a collision, and they do so instantaneously. This view is appropriate when the thermal energy is much larger than the energy splitting between states.
 In order to spin up the water molecule, an impactor must collide with one of the hydrogen atoms. The impactor is an atom within another water molecule, with velocity $v$ relative to the molecule to be hit. The colliding particle in the impactor is most likely to be the (larger) oxygen atom. After collision the velocity of the hit hydrogen atom is
\begin{equation}
v_H \simeq \frac{ m_O v}{m_H + m_O} \simeq v.
\end{equation}
If the collider were a hydrogen, $v_H$ would be halved. The rotation frequency of the impacted molecule about the massive oxygen atom is $\omega = v_H/L$ where $L$ is the bond length. The amount of energy in rotation is
\begin{equation}
\frac{1}{2} I_W \omega^2 \simeq \frac{ I_W  v^2}{2 L^2}\, ,
\end{equation}
where $I_W$ is the moment of inertia of the water molecule. The ratio of rotational energy to incoming translational energy is given by
\begin{equation}
\frac{E_{\rm rot}}{E_{\rm trans}} \simeq \frac{ I_W}{m_W L^2} \, .
\end{equation}
Using the mean moment of inertia $\overline{I_W} = 2 \times 10^{-40} {\rm ~g~cm}^2$ and the bond length $L = 1.0$ \AA$^2$, we obtain
\begin{equation}
\frac{E_{\rm rot}}{E_{\rm trans}} \simeq 0.05
\end{equation}
for the fraction of energy transferred into molecular rotation after a single collision. The fractional energy loss to rotation $\xi = 0.05$ is thus substantially less than the limiting value for heating derived in Section \ref{sec:stir}, $\xi_{\rm lim} = 3/11 \simeq 0.27$.  Heating can occur. 

We adopt the fractional inelasticity parameter derived here  $\xi = 0.05$ for the remainder of the paper.

\section{Spreading and heating of the torus}
\label{sec:spread}

We have established so far that the collision rate in the cold torus is conducive to rapid heating, and that rotational line cooling is not sufficient to prevent this heating. As the torus heats and viscously spreads, the density will fall and the collision cross section will change. To improve upon our simple estimates we take these factors into account and make a simple model for the time evolution of the water torus, assuming it begins unheated and lasts for the charge exchange time, i.e. the torus contains $N_{\rm tot} = \dot{N} \tau_{\rm ce}$, where the source rate is believed to be $\dot N \simeq 1 \times 10^{28} \rm s^{-1}$.

We again adopt axisymmetry and assume an initial temperature 180K (i.e. $u(0)=0.5 {\rm ~km~s}^{-1}$).   We solve the following simplified differential equations. The number density of water molecules in the torus centered around $4 R_S$ is
\begin{equation}
n(t) = \frac{N_{\rm tot}}{2 \pi (4 R_S) W_r(t) W_z(t)} \, ,
\end{equation}
where $W_r$ is the full radial width of the torus, and $W_z$ is the full vertical width. Viscous spreading occurs at speed $\sim \nu/\Delta R$, where $\Delta R$ is the lengthscale over which the angular momentum varies. Viewing the torus as two segments each of lengthscale $W_r/2$, the radial spreading rate is approximated as 
\begin{equation}
\frac{dW_r}{dt} = \frac{4 \nu(t)}{W_r(t)},
\end{equation}
where the viscosity $\nu$ is obtained from equation (\ref{eq:visc}) with $\xi=0.05$. The velocity dispersion $u^2$ evolves according to
\begin{equation}
\frac{d u^2}{dt} = u^2(t) \left(\frac{\zeta}{\tau_{\rm coll}(t)}+\frac{2}{3} \frac{dn/dt}{n(t)}\right),
\end{equation}
where the second term in brackets represents cooling due to expansion of the torus, approximating it as a gas with adiabatic index $5/3$.
We also employ
\begin{equation}
W_z(t)=\frac{2}{\Omega_K}\sqrt{\frac{8u^2(t)}{33}}
\end{equation}
for the full vertical width of the torus (twice the scale height), where we used Eq. \ref{eq:uz} to obtain the vertical velocity dispersion. The collision rate is 
\begin{equation}
\frac{1}{\tau_{\rm coll}(t)} = n(t) \sigma_{\rm coll}(t) u(t) \, ,
\label{eq:colltime}
\end{equation}
where $\sigma_{\rm coll}$ comes from Eq. \ref{eq:sigmaT}.
Using Eq. \ref{eq:ur} for the radial velocity dispersion, the torus has initial radial width 
\begin{equation}
W_r(0)=\frac{2}{\Omega_K}\sqrt{\frac{20u^2(0)}{33}} \, .
\end{equation}

\begin{figure}[htbp]
\begin{center}
\includegraphics[width=3in]{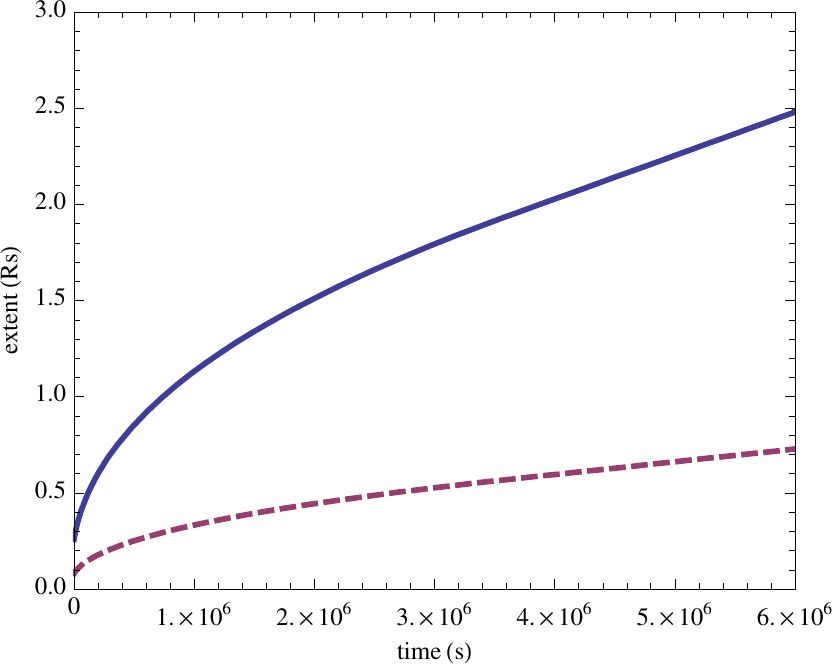}
\caption{Initial spreading and heating of the Enceladus torus, for $u(0)=0.5{\rm ~km~s}^{-1}$. Solid line: radial width of torus; dashed line: vertical scale height.}
\label{fig:heat}
\end{center}
\end{figure}

Figure \ref{fig:heat} shows the result of this calculation. Although the density and collision rate fall, the torus continues to heat. After the charge exchange time, the radial extent of the torus is $W_r \simeq 2.5 R_S$ and the vertical scale height is of order $W_z \simeq 0.7 R_S \simeq 0.18 R$. Because now $W_r \sim R$ the neglect of radial variations is no longer valid; we address this in the next Section.

\begin{figure}[htbp]
\begin{center}
\includegraphics[width=3in]{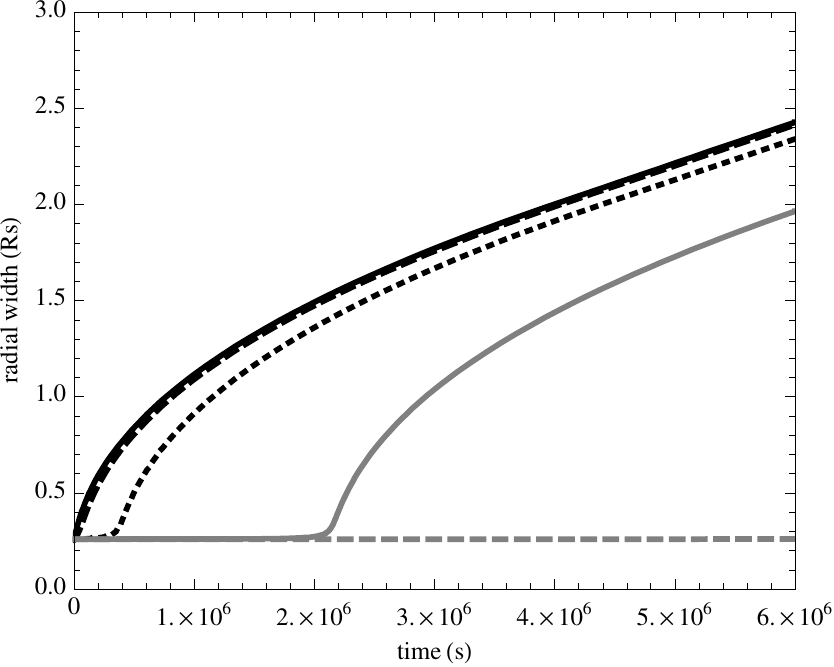}
\caption{Evolution of radial width of torus, with variation in initial temperature. From top to bottom, initial velocity dispersion $u=0.5,0.3,0.1,0.03,0.01 {\rm~km~s}^{-1}$.}
\label{fig:temps}
\end{center}
\end{figure}

We also investigate the effect of varying the initial temperature of the water in the torus. At lower temperatures, the torus is initially more dense, and the collision rate may be larger than the orbital frequency. In these cases we use the following relation for the viscosity
\begin{equation}
\frac{1}{\nu^2} = \frac{1}{\nu_1^2}+\frac{1}{\nu_2^2} \, ,
\end{equation}
where $\nu_1$ is the high density-regime viscosity from Eq. \ref{eq:high_visc}, and $\nu_2$ comes from the low-collision rate expression, Eq. \ref{eq:visc}. In the high density case, the viscous heating rate is small and is set equal to $du^2/dt= 2 \nu_1 (3\Omega_K/2)^2$.

Figure \ref{fig:temps} displays the results of this calculation. When $u(0) 
\gtrsim 0.1 {\rm ~km~s}^{-1}$ the width after one charge exchange time is similar to the case when $u(0) = 0.5 {\rm ~km~s}^{-1}$. For smaller initial velocity dispersions, the torus remains in the high collision-rate regime and never reaches the critical point $\Omega_K \tau_{\rm coll} \simeq 1$ where the viscosity is largest. The spreading is therefore much less.

\section{Steady-state water torus}
\label{sec:sstate}

We now adopt the results of the previous section for initial temperature 180K, i.e.
\begin{equation}u_z = 0.18\, \Omega_K R \; ; \; u_r = 0.28\, \Omega_K R \, ,
\label{eq:vels}
\end{equation}
then calculate the expected  steady-state radial distribution of water in the torus. In steady state the torus, surface mass density $\Sigma$, evolves according to
\begin{equation}
\frac{\partial \Sigma}{\partial t} = 0 = \frac{3}{R} \frac{d}{dR}\left(R^{1/2} \frac{d}{dR}\left[\nu(R) \Sigma R^{1/2}\right]\right)-\frac{\Sigma}{\tau_{\rm loss}(R)} \, ,
\label{eq:sstate}
\end{equation}
where the first term on the right is the standard equation for radial viscous spreading in a thin Keplerian disk \citep{frank.1992} and the second is the loss rate of neutrals to charge exchange and ionization. Strictly speaking the above equation is valid only  when $u \ll \Omega_K R$, i.e. when local disk properties can be related to local values of transport coefficients. Clearly we are close to the edge of validity of this approximation. In particular, it neglects the high-energy, high-eccentricity tail of the particle distribution, and as such we expect the amount of spreading calculated to be a lower limit.

From Eq. \ref{eq:visc} and \ref{eq:colltime}, the viscosity is given by 
\begin{equation}
\nu(R) = \frac{1}{12.6} \frac{u^3(R) n(R) \sigma_{\rm coll}(R)}{\Omega_K^2(R)}.
\end{equation}
We convert to surface mass density by means of
\begin{equation}
\Sigma(R) = m_{\rm H_2O} n(R) \frac{u_z(R)}{\Omega_K(R)},
\end{equation}
and for the collision cross section we use Eq. \ref{eq:sigmaT} expressed in terms of velocity dispersion $u$:
\begin{equation}
\sigma_{\rm coll} = 80 \left(\frac{u}{0.5 {\rm~km~s}^{-1}}\right)^{-4/3} {\rm ~\AA}^2 \, .
\label{eq:sigmav}
\end{equation}
Augmenting Eqs. (\ref{eq:vels})-(\ref{eq:sigmav}) with Kepler's law $\Omega_K \propto R^{-3/2}$ and scaling all quantities to the orbit of Enceladus, we obtain
\begin{equation}
\nu \simeq 0.29 \left(\frac{R}{R_0}\right)^{7/6}\left( \frac{\Sigma/m_{\rm H_2O}}{\rm cm^{-2}}\right) {\rm ~cm^2~s^{-1}} \, ,
\end{equation}
where $R_0=4 R_S$.

We then solve the differential equation (\ref{eq:sstate}) numerically, setting the surface density equal to zero at the edge of the rings, i.e. the rings are considered perfect absorbers. We place the edge of the rings at an effective radius larger than their actual edge at $R_{\rm edge} \simeq 2.2 R_S$: an eccentric orbit centered at distance $R$ reaches a minimum radius $R(1-e)$ at periapse, and if the orbit anywhere intersects the rings the particle will be absorbed. Because we have assigned an eccentricity $e\simeq 0.28$ to all particles, the edge of the rings is placed at $R_{\rm eff} = R_{\rm edge}/(1-e) \simeq 3.0 R_S$.

As the torus spreads and heats, the neutral water molecules spend less time in the regions of high ion density, and so the lifetime to charge exchange increases. The rate of photodissociation to OH+H, however, is independent of radius and is similar to the charge exchange rate near Enceladus (see Table \ref{tab:proc}). We consider two cases. First, we keep $\tau_{\rm loss}$ constant with radius and equal to the photodissociation time. This corresponds to treating dissociation as a loss mechanism. Second is the more physically realistic case, in which we retain the OH and treat it in the same way as water: OH has a similar dipole moment \citep[see e.g.][]{sauval.1984} and mass to water, so its neutral interactions are similar. We neglect any energy gained by the OH upon dissociation, since we expect most of this to go to the liberated hydrogen atom. In this second case, we adopt the following formula for the loss timescale as a function of radius
\begin{equation}
\tau_{\rm loss} = \left[1+0.5 \left(\frac{R}{6 R_S}\right)^5\right]  6 \times 10^6 {\rm ~s},
\end{equation}
up to a maximum of $1.8 \times 10^8$ s, corresponding to the photoionization timescale. This expression was obtained as an approximate fit to fig. 17 of \cite{sittler.2008}, who calculate the neutral loss timescale as a function of radius, including loss to charge exchange, electron impact ionization and solar photoionization.

\begin{figure}[htbp]
\begin{center}
\includegraphics[width=3in]{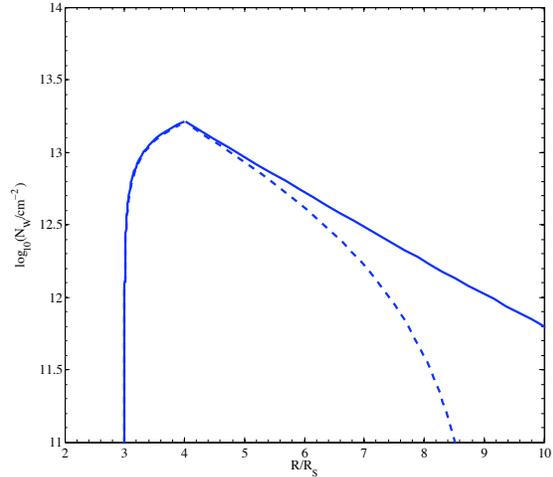}
\caption{Calculated column density profile of the Enceladus water torus. Dashed line: treating photodissociation as a loss mechanism. Solid line: treating only ionization and charge exchange as loss mechanisms; line represents water + OH column density in this case.}
\label{fig:coldens}
\end{center}
\end{figure}

The neutral column density for both cases is plotted in Fig. \ref{fig:coldens}. Both profiles are in reasonable agreement with the observed OH column density from \cite{jurac.2002,jurac.2005}. The true OH column is bracketed by the two lines plotted. It must be noted that a direct comparison between these curves and the measured OH column is inappropriate for the following reasons: first, we have plotted water or water+OH, not OH alone; and second, we have neglected spreading of the torus via the other established mechanism, charge exchange \citep{johnson.2006}. 

It is of interest to compare the derived full width of the torus with the result of the calculation in Section \ref{sec:spread}. From Fig. \ref{fig:coldens}, the full width between points with $N_W = N_{\rm W,max}/e$  is $\Delta R \sim 2.7 R_S$, consistent with $W_r \simeq 2.5 R_S$ derived in Section \ref{sec:spread}.

\section{Implications and comparison with previous work}
\label{sec:discuss}

Our results serve to show that significant spreading of the torus can occur via neutral-neutral interactions alone, and thus such interactions, with correct cross sections, should be in future included in the global models such as those of \cite{jurac.2005} which include all relevant processes.

A further issue that can be addressed is deposition of water on to the outer parts of the rings. \cite{jurac.2007} estimate using their model that 17\% of the water from Enceladus is precipitated on to the rings. \cite{farrell.2008} report measurements of a decrease in plasma density at the edge of the rings, and use this to calculate the rate of absorption of ions. In both of the cases shown in Fig. \ref{fig:coldens}, about 70 kg~s$^{-1}$ (25\%) of the total 280 kg~s$^{-1}$ released by Enceladus precipitates on to the rings. This is probably a lower limit, both because other processes also contribute to loss on to the rings; and because our model probably represents a lower limit to the degree of spreading of the torus.

The detailed models of \cite{jurac.2002,jurac.2005,jurac.2007} do not show significant viscous heating or interactions between neutrals. We suggest that this is because their adopted water-water cross section was that of atomic oxygen \citep{jurac.2005}, which has radius $r \simeq 0.6$ \AA. In fact, the characteristic collisional size of the molecule is substantially larger, at least that of the O$^{2-}$ ion ($r \simeq 1.4$ \AA), and is further enhanced by the permanent dipole moment as detailed in the Introduction. Because there are fewer collisions between neutrals before they are impacted by other processes, they do not spread or heat significantly in the models of Jurac et al.

\cite{marconi.2003} performed monte carlo simulations of the low-collision rate atomic hydrogen torus around Neptune's moon Triton. This approach follows individual particles and so does not suffer the potential inaccuracies of our thin disk model. The comparison between Marconi's results and ours, however, is encouraging. After a similar number of collision times, the torus has spread by only a slightly larger amount than we find in our Section \ref{sec:sstate}. 

\section{Observing the water torus remotely}
\label{sec:obs}

The low density of the extended water torus makes it difficult to observe in situ and remotely. Only in the fluorescent emission of the hydroxyl radicals has the extended neutral torus been seen \citep{shemansky.1993,jurac.2002}, while only very close to Enceladus are the neutral densities large enough to be measured in situ \citep{waite.2006}. Once ionized, detection is easier, but the additional electromagnetic forces felt by ions and electrons renders the inference of the distribution of their parent neutral population difficult, although this has been attempted \citep{sittler.2008}. The majority of the extended neutral water cloud has never been directly detected. The upcoming \textit{Herschel} mission has the ability to change this.

One of the science goals of the \textit{Herschel Space Observatory}, due for launch in 2008, is the observation of water rotational lines in the nearby universe \citep[e.g.][]{degraauw.2006} with unparalleled sensitivity and angular resolution. The HIFI instrument on board Herschel provides continuous frequency coverage from 480 to 1250 GHz, covering $\sim 10$ of the lower rotational transitions of the water molecule, at frequency resolutions up to $\nu/\Delta \nu =10^7$ \citep{degraauw.2006}. The 3.5 m diameter of Herschel provides angular resolution at 560 GHz of $\sim 0.5$ arcmin, corresponding to $\sim 3.6 R_S$ at the distance of Saturn. This resolution is sufficient to crudely map the neutral torus.

We now estimate the signal strength from the torus. For definiteness, we consider the $1_{10} -  1_{01}$ (ortho) transition at 560 GHz, and use our steady-state hot torus model from the previous section. We take the optimistic view that every collision leads to an excitation of this line (but neglect emission due to electron collisional and solar excitation, which will be significant), and calculate the flux per beam of \textit{Herschel} as a function of distance along the equator, where Saturn is assumed to be at equinox.

The received flux is expressed in the standard way in terms of an antenna temperature $T_{\rm ant}$:
\begin{equation}
k T_{\rm ant} = \frac{\lambda^2 B_\nu}{2} \, ,
\end{equation}
where $B_\nu$ is the specific intensity of the source \citep[e.g.][]{burke.2001}.
Emission in the 560 GHz line is spread by both thermal and orbital Doppler shifts; we adopt a linewidth characteristic of the local velocity dispersion, which is a fraction of the orbital velocity in our model, $\Delta \nu = v_{\rm disp}(R)/c$. The typical linewidth is then of order $\Delta \nu \simeq 20$ MHz, i.e. $\nu/\Delta \nu \simeq 3 \times 10^4$.

\begin{figure}[htbp]
\begin{center}
\includegraphics[width=3in]{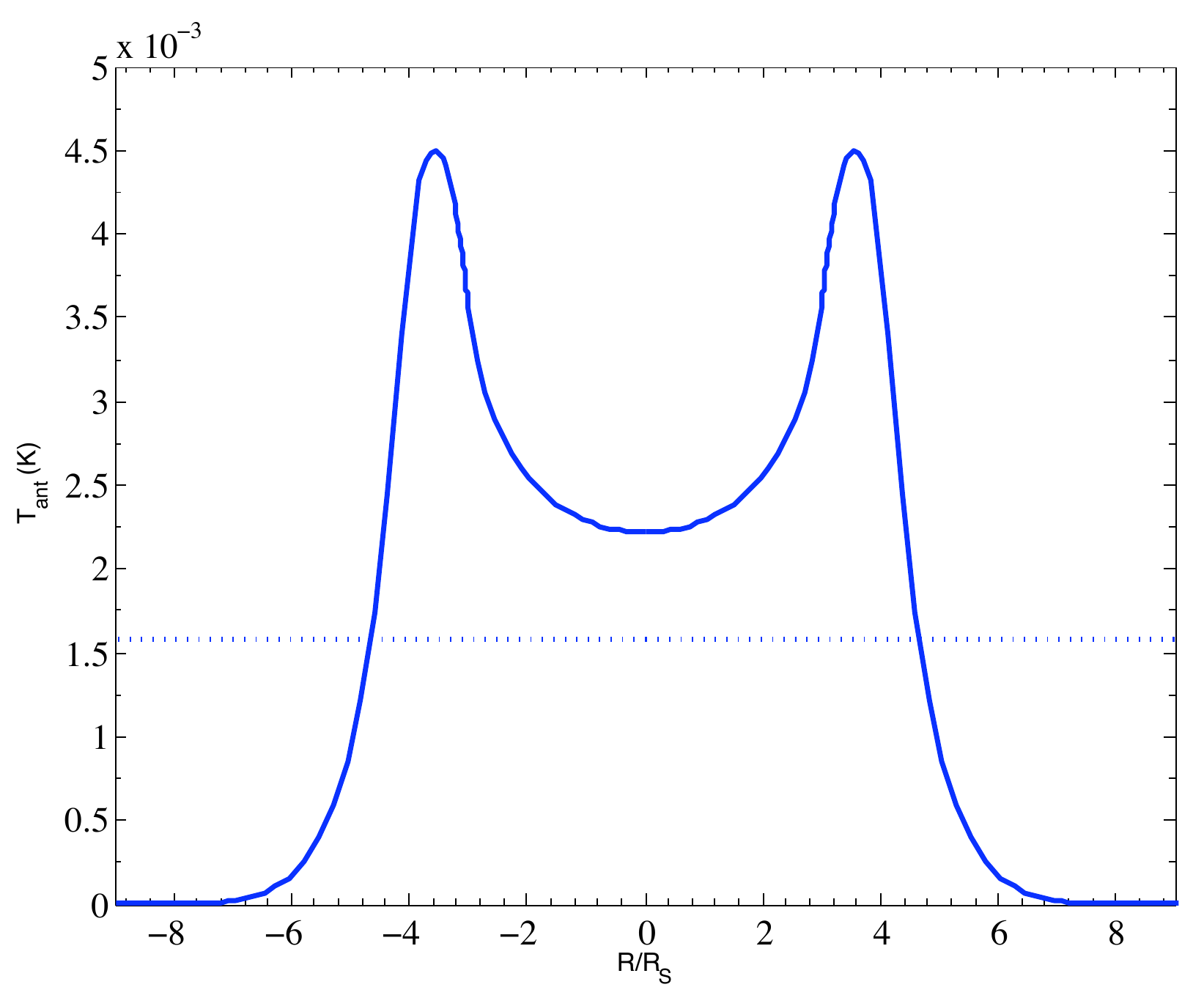}
\caption{Antenna temperature in the lowest ortho-water rotational line, with beam centered on equator. Horizontal axis shows distance along equator from center of planet. Signal integrated through cloud assuming it is optically thin. Horizontal dotted line shows 5-sigma detection limit for \textit{Herschel} HIFI instrument after 1 hour of integration at frequency resolution $\nu/\Delta\nu = 10^4$.}
\label{fig:herschel}
\end{center}
\end{figure}

Integrating along lines of sight through the equatorial plane of the torus, and neglecting beam dilution due to vertical variation in density, we obtain the antenna temperature profile shown in Fig. \ref{fig:temps}. With a system temperature $T_{\rm sys} = 160$ K in the lowest band of the HIFI instrument, Herschel has a 5-sigma detection limit of 1.6 mK, in 1 hour of observation at frequency resolution $10^4$ \citep{degraauw.2006}, and this is also plotted in Fig. \ref{fig:herschel} as the horizontal dotted line.

Under the assumed conditions, then, the neutral water torus is readily observable by \textit{Herschel}. The angular resolution prevents the detailed examination of the spatial distribution, but by comparing the flux in various lines it should be possible to determine the collision rate in the torus and to gauge the temperature of the water via linewidths. A more comprehensive model of the torus should be used to make detailed predictions for different physical models.

\subsection{Absorption by the torus}

In the above the torus was assumed optically thin in the 560 GHz line. In fact the optical depth is close to unity and a bright background source could allow observation of the torus to further distances from the planet, where less emission occurs.

The optical depth in the line is given by \citep[e.g.][]{rybicki.1979}
\begin{equation}
\tau_{\rm line} \simeq  \frac{N_W \lambda^2}{8 \pi} \frac{A_{1_{10} -  1_{01}}}{\Delta \nu},
\end{equation}
where $\Delta \nu$ is the linewidth and $N_W$ is the line of sight column density in water molecules in the lower of the two rotational states (the vast majority of molecules, as discussed in Section \ref{sec:cooling}). Using the spontaneous emission coefficient from Table \ref{tab:rot} and the line of sight column density at 6 $R_S$ from our steady-state model in Section \ref{sec:sstate}, we obtain
\begin{equation}
\tau_{\rm line} \simeq 1 \, .
\end{equation}
The torus is thus readily detectable in absorption against any sufficiently bright millimeter source that passes behind it. It small angular size, however, makes the existence of any such source rather unlikely; the cosmic microwave background could be used, but because of its near isotropy on the sky, the strength of the signal would be reduced by a factor $\sim v_{\rm disp}/c$.

It is clear from this work that neutral-neutral interactions in the Enceladus torus cannot be neglected; they alone can account for the observed radial extent of the hydroxyl torus, though undoubtedly spreading also occurs through charge exchange reactions. The water-water collisions, with the cross sections calculated here, should be included in future global simulations of the torus. This may impact our estimates of ice deposition rates on the rings, as well as the required source rate to sustain the observed torus. The \textit{Herschel} mission will provide a timely means of observing the neutral water for the first time, by means of its rotational lines, and may be able to distinguish between alternative neutral cloud models.

\acknowledgments
The author wishes to thank R. Narayan and P. Goldreich for productive conversations.

\end{document}